\documentclass[USenglish,twocolumn]{article}

\usepackage[utf8]{inputenc}

\usepackage[big]{dgruyter}
\usepackage{microtype}

\usepackage{amssymb}
\usepackage[export]{adjustbox}
\usepackage{amsmath}
\usepackage{graphicx}
\usepackage[multi-part-units=single]{siunitx}
\usepackage{url}
\usepackage{multirow}
\usepackage{svg}
\usepackage{amsmath}
\usepackage{siunitx}
\usepackage{float}
\usepackage{tikz}

\begin{document}

  \articletype{Research Article}

  \author[1]{M. Bengs$^{1}$}
  \author[1]{N. Gessert$^{1}$}
  \author[1]{A. Schlaefer}
    
  \runningauthor{M. Bengs et al.}
  \affil[1]{Institute of Medical Technology, Hamburg University of Technology, Hamburg, Germany, \newline E-mail: marcel.bengs@tuhh.de \newline $^{1}$ Both authors contributed equally.}

  \title{4D Spatio-Temporal Convolutional Networks for
Object Position Estimation in OCT Volumes}
  \runningtitle{4D Spatio-Temporal Convolutional Networks for
Object Position Estimation in OCT Volumes}
  \subtitle{...}
  
  \abstract{Tracking and localizing objects is a central problem in computer-assisted surgery. Optical coherence tomography (OCT) can be employed as an optical tracking system, due to its high spatial and temporal resolution. Recently, 3D convolutional neural networks (CNNs) have shown promising performance for pose estimation of a marker object using single volumetric OCT images.  While this approach relied on spatial information only, OCT allows for a temporal stream of OCT image volumes capturing the motion of an object at high volumes rates. In this work, we systematically extend 3D CNNs to 4D spatio-temporal CNNs to evaluate the impact of additional temporal information for marker object tracking. Across various architectures, our results demonstrate that using a stream of OCT volumes and employing 4D spatio-temporal convolutions leads to a 30\% lower mean absolute error compared to single volume processing with 3D CNNs.}
  
  \keywords{Convolutional Neural Networks, Spatio-temporal data, Position Estimation, Optical Coherence Tomography}
  \classification[PACS]{...}
  \journalname{Current Directions in Biomedical Engineering}
  \journalyear{2020}
  \journalvolume{}
  \journalissue{}
  \startpage{1}
  \aop

\maketitle

\section{Introduction}
Minimally invasive surgery (MIS) enables fewer post-operative complications compared to open surgery, by significantly reducing the access incisions and surgery trauma \cite{dogangil2010review}. However, performing MIS is a challenging task, due to a limited field of view and lacking perception of force feedback, which requires computer-assisted surgery, particularly precise surgery tool tracking. In this regard, several vision-based approaches using images and videos have been proposed \cite{Bouget2017}. While 2D images and videos only provide 2D spatial information, typical tissue structures and object movements are inherently three dimensional. Therefore, for many medical applications using volumetric imaging is preferable or required, e.g. for prostate radiation therapy \cite{chinnaiyan20033d}, or for precise pose estimation of a marker object \cite{Gessert2018}. Some modalities provide not only volumetric images, but also allow for imaging with a high temporal resolution, such as optical coherence tomography (OCT), and hence can be used as an imaging modality for an optical tracking system \cite{schluter2019feasibility, laves2017feature}. \\
To overcome the limitations of classical tracking approaches, which rely on handcrafted features limited to specific application scenarios such as skin \cite{laves2017feature} or eye motion tracking \cite{camino2016evaluation}, deep learning has been proposed recently. In particular, 3D convolutional neural networks (CNNs) have shown promising results for precisely localizing small objects based on OCT-data \cite{Gessert2018}. This approach employed 3D CNNs on a single volumetric image, allowing to turn arbitrary small objects into a marker for pose estimation. However, as OCT allows for a temporal stream of OCT image volumes, it seems reasonable that the preceding image volumes at high volume rates may carry information on the object's motion. This leads to the challenging problem of 4D deep learning, which is largely unexplored so far and has only been addressed in a few applications such as functional magnetic resonance imaging \cite{bengs2019a}, computed tomography \cite{clark2019convolutional} and OCT-based force estimation \cite{gessert2020deep} as well as OCT-based tissue motion estimation \cite{bengs2020spatio}.\\
\begin{figure*}
\centering
\includegraphics[width=0.60\textwidth]{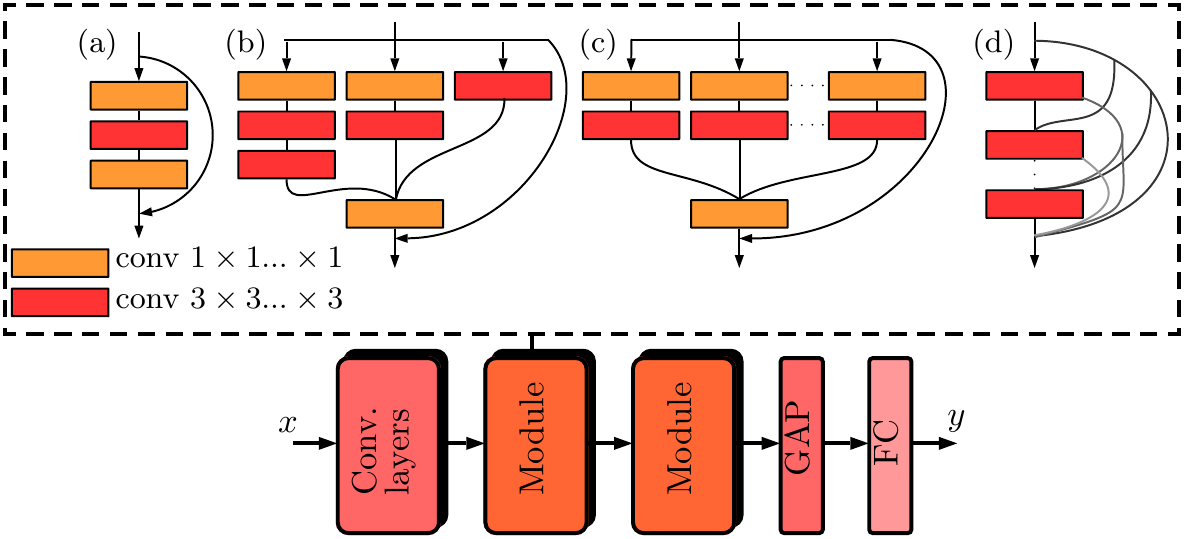}%
\caption{Each of our custom architecture consists of an initial part with 5 convolutional layers, followed by architecture modules that represent subsequent architecture blocks. Note, the first block in each module downsamples the input dimensions by a factor of two. The different architecture blocks are (a) ResNet, (b) Inception (c), ResNeXt, and (d) Densenet. We use a global average pooling (GAP) layer after the last module, and the output is directly fed into a fully connected output layer (FC).
}
\label{fig:modules}
\end{figure*} In this work, we systematically extend 3D CNNs to 4D spatio-temporal data processing and evaluate whether a stream of OCT volumes improves object position estimation performance. Spatio-temporal processing with CNNs can be done by stacking multiple frames into the channel dimension \cite{Pfister2014}, or by using full or factorized spatio-temporal convolutions \cite{Tran2015,Qiu2017}. Even though these methods have shown promising performance for video analysis tasks \cite{Tran2015,Qiu2017,Pfister2014}, it is largely unclear how CNNs perform with 4D, as they have not been systematically studied. Therefore, we evaluate four widely used CNN architectures and consider several different types of convolutions for 4D data processing. We employ volume stacking, factorized, and full spatio-temporal convolutions, and compare the position estimation performance to single volume processing. For systematic evaluation of our methods, we consider the problem of position estimation of a marker object, with a specialized OCT setup which enables fast acquisition of sufficient 4D data with a well-defined ground-truth. 

\section{Materials and Methods} 
\subsection{Network Architectures} 

\begin{figure}
\centering
\includegraphics[width=0.35\textwidth]{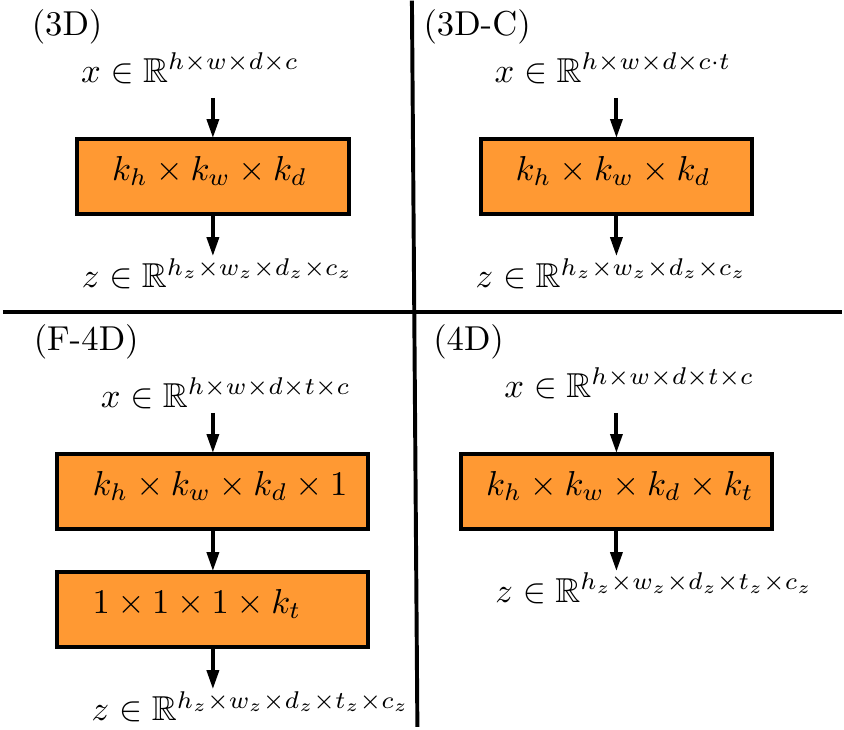}%
\caption{The different convolutions we employ: (3D) 3D spatial convolution; (3D-C) 3D convolution, with temporal information stacked into the channel dimension; (F-4D) Factorized 4D spatio-temporal convolution; (4D) 4D spatio-temporal convolution.}
\label{fig:conv_op}
\end{figure}

We evaluate four different methods with four different architectures to predict the current position of a marker object using a stream of OCT volumes. Similar to a previous approach \cite{Gessert2018}, we define our own architectures following the architecture principles of four widely used state-of-the-art architectures, ResNet, Inception, ResNeXt, and Densenet. Each of our custom architecture consists of an initial part with five convolutional layers, followed by architecture modules, shown in Figure \ref{fig:modules}. Note, the number of building blocks inside the modules are tuned based on validation performance. For each architecture, we evaluate four different types of convolutions, see Figure \ref{fig:conv_op}. \\ 
First, we consider a previous approach on marker object tracking \cite{Gessert2018}, and use 3D convolutions applied to single volumetric images, which is our baseline. (3D)
\\ 
Second, we stack multiple consecutive volumes into the channel dimension of the network's input and use a 3D convolution.  (3D-C)
\\ 
Third, we examine factorized spatio-temporal convolutions \cite{Qiu2017}, which split a full spatio-temporal convolution into a temporal and a spatial convolution. Every single spatio-temporal convolution is replaced by two successive factorized 4D convolutions. Note, there are no native implementation of 4D operations available for standard libraries such as Tensorflow or PyTorch. Hence, we implement our custom 4D convolution and pooling operation in Tensorflow, using multiple native 3D convolution and pooling operations across multiple time-shifted volumes. (F-4D)
\\ 
Fourth, we consider 4D spatio-temporal convolutions and replace each 3D convolution and 3D pooling with the corresponding 4D counterparts. (4D)
\\ 
The networks are trained for 350 epochs with a batch size of 18 and Adam for optimization of the mean squared error (MSE) loss function.

\subsection{Data Set} 
\begin{figure*}
	\centering
\includegraphics[height=2.20cm]{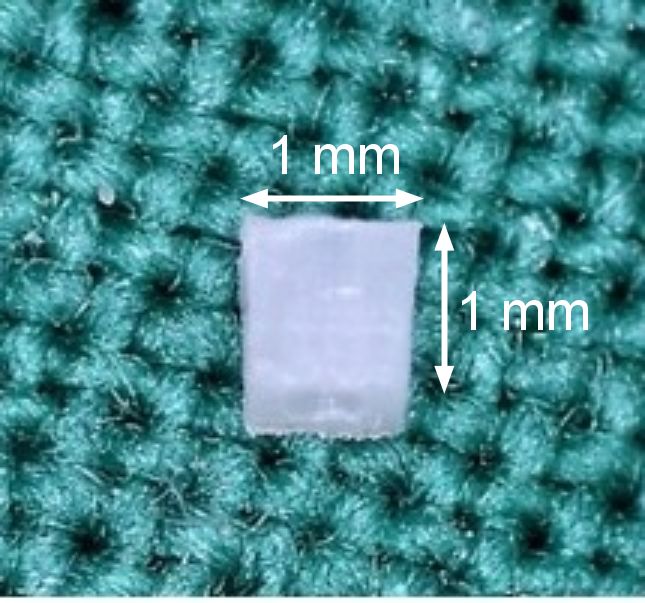}\hspace{0.01cm}	
\begin{tikzpicture}
\node[anchor=south west,inner sep=0] (image) at (0,0) {\includegraphics[height=2.20cm]{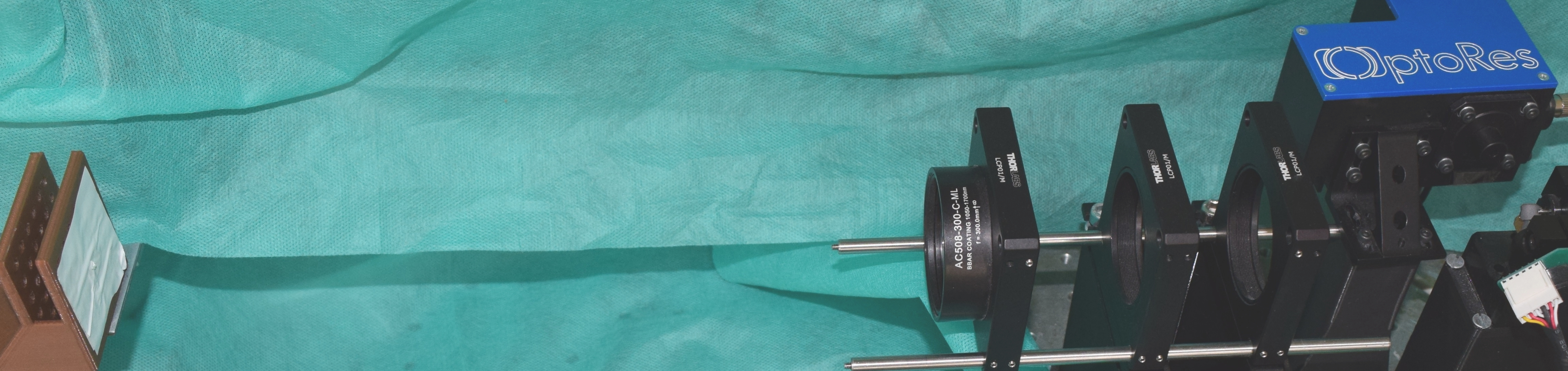}};
\node [anchor=west, white] (marker) at (1.0,1.5) {Marker};
\node [anchor=west, white] (OCT) at (5,2) {OCT};
\node [anchor=west, white] (galvos) at (7,0.5) {Galvos}; 
\node [anchor=west, white] (lens) at (2,0.5) {Achromatic lens};
  
   \begin{scope}[x={(image.south east)},y={(image.north west)}]
      
        \draw [->, line width=1pt, white] (marker) --(0.06,0.300);
        \draw [->, line width=1pt, white] (OCT) --(0.896,0.9036);
        \draw [->, line width=1pt, white] (galvos) --(0.971,0.4209);
        \draw [->, line width=1pt, white] (lens) --(0.585,0.3845);                        
    \end{scope}
\end{tikzpicture}%
	\caption{The experimental setup: Marker object (left); OCT setup (right). The marker object is attached to a holder.}
	\label{fig:Experimental-Setup}
\end{figure*}
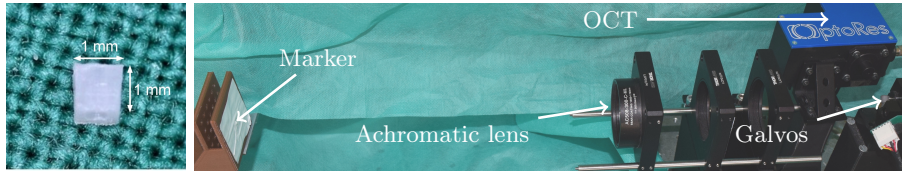

For data acquisition, we use a commercially available swept-source OCT device (OMES, OptoRes), a second scanning stage with two mirror galvanometers, an achromatic lens, a marker object, and a holder for the marker object. The marker object is made of a polyoxymethylene block with a size of $1\mathrm{\,mm^3}$. The whole setup is shown in Figure \ref{fig:Experimental-Setup}. We consider volumes with a size of $32\times32\times32$ with a corresponding field of view (FOV) of $3\,\mathrm{mm}\times3\mathrm{\,mm}\times3.5\mathrm{\,mm}$, and an acquisition speed of 833 volumes per second. Our OCT setup is enhanced with a second scanning stage with two mirror galvanometers controlled by stepper motors, which enable to shift the FOV in the lateral dimensions. Also, a third motor shifts the FOV in the axial dimension, by setting the pathlength of the OCT's reference arm. In this way, our OCT-setup allows for shifting the FOV in all spatial directions without moving the scan head. This can be utilized for automatic OCT volume acquisition and ground-truth annotation. In particular, instead of moving the marker object, we move the FOV of the OCT and the current motor positions represent the relative marker position in the FOV. \\ Next,  we repeat the following steps and define a set of target motor positions that shift the FOV, representing smooth marker movements.  First, a set of 60 to 90 target positions $n_{j}$ are randomly generated for the three stepper motors. Then, piecewise cubic spline interpolation $f:\mathbb{R}_{+}\rightarrow\mathbb{R}^{3},\,\,\tau\mapsto f(\tau)$ is used to obtain a smooth function connecting the  target positions, $f(\tau_{j})=n_{j}$. Afterwards, 500 motor points are sampled from the piecewise cubic spline function $f(\tau)$ with equidistant parameter values $\tau$. Note, this does not lead to equidistant data points, due to the curvature of the spline function.  We repeat this procedure, to obtain the full data set with 7000 examples.  Afterward, we acquire one volumetric image for each target motor positions, that serve directly as ground truth annotation.  In summary, we use 5000 volumes for training and 1000 each for validating and testing our models. For our experiments, we evaluate a sequences length of five consecutive volumes. The corresponding target $t\in\mathbb{R}^{3}$ refers to the last position in one sequence. 

\section{Results}
We report the mean absolute error (MAE) and relative mean absolute error (rMAE) for our experiments in Table \ref{tab:All-networks-with metrics}. The MAE is given in \SI{}{\micro\metre} based on a calibration between galvo motor steps and image coordinates. The rMAE is relative to the target's standard deviation. Overall, using temporal data improves performance for all architectures, while 4D spatio-temporal convolutions perform best. On average the inference times are 6 ms and 20 ms for 3D and 4D architectures, respectively.

\section{Discussion and Conclusion}

\begin{table}
\caption{Comparison of the different architectures with the different types of convolutions.\label{tab:All-networks-with metrics}}
\begin{center}
\begin{tabular}{l l l l l l}
 & Type & MAE (\SI{}{\micro\metre}) & rMAE &  Parameters \\
\hline 
\multirow{4}{*}{\rotatebox[origin=c]{90}{ResNet}}

 & 3D & $15.87\pm14.40$ & $0.013\pm0.011$ &  \num{409755} \\
 & 3D-C & $13.39\pm10.96$ & $0.011\pm0.009$ &  \num{411483}\\
 & F-4D & $12.36\pm10.15$ & $0.010\pm0.008$ &  \num{454575} \\
 & \textbf{4D} & $\mathbf{11.79\pm9.79}$ & $\mathbf{0.009\pm0.008}$ & \num{1137459} \\

\hline 
\multirow{4}{*}{\rotatebox[origin=c]{90}{Inception}} 
 & 3D & $17.65\pm15.48$ & $0.014\pm0.012$  & \num{428521} \\
 & 3D-C & $14.83\pm11.84$ & $0.012\pm0.009$ & \num{430249}\\
 & F-4D & $13.23\pm11.36$ & $0.010\pm0.009$  & \num{475006} \\
 & \textbf{4D} & $\mathbf{11.87\pm9.66}$ & $\mathbf{0.009\pm0.008}$ & \num{1161568} \\ 

 \hline 
\multirow{4}{*}{\rotatebox[origin=c]{90}{ResNeXt}}
 & 3D & $16.96\pm15.53$ & $0.013\pm0.012$  & \num{392367} \\
 & 3D-C & $13.00\pm12.16$ & $0.010\pm0.010$  & \num{394095}\\
 & F-4D & $12.32\pm10.99$ & $0.010\pm0.009$  & \num{432903} \\
 & \textbf{4D} & $\mathbf{11.87\pm10.93}$ & $\mathbf{0.009\pm0.009}$ & \num{1012215} \\
 
 \hline 
\multirow{4}{*}{\rotatebox[origin=c]{90}{Densenet}}
 & 3D & $16.03\pm13.69$ & $0.013\pm0.011$  & \num{406723} \\
 & 3D-C & $14.39\pm11.57$ & $0.011\pm0.009$  & \num{445139}\\
 & F-4D & $12.51\pm10.07$ & $0.010\pm0.008$  & \num{420205} \\
 & \textbf{4D} & $\mathbf{11.54\pm9.51}$ & $\mathbf{0.009\pm0.008}$ & \num{1080683}\\

\end{tabular}
\end{center}
\end{table}

Our results in Table 1 show that using a sequence of volumes consistently outperforms single volume usage. This agrees with our expectation that a temporal stream of volumetric images should improve position estimates. Analyzing the different types of temporal processing shows that increasing complexity of the 4D image processing results in better predictions. Stacking the volume sequence in the channel dimension already improves performance by 15\% on average compared to using a single volumetric input, while the number of parameters remain similar. This indicates that even with processing only at the network's input, valuable temporal information can be extracted. Note that temporal information is lost after the first convolution operation, because no temporal convolutional operation is performed \cite{Tran2015}. Using 4D factorized convolutions instead improves performance by 25\% on average compared to using a single volumetric input, and the number of parameters is only increased by less than 11\%.  This shows that the 4D data structure can be leveraged by factorized convolutions similar to previous findings on 3D spatio-temporal data \cite{Qiu2017}. Finally, full 4D spatio-temporal convolutions lead to the best performance, demonstrating that 4D CNNs are able to extract valuable spatio-temporal features from 4D data.\\ Moreover, our methods perform consistently across different network architectures. Notably, the type of network architecture only has a minor impact on the errors, while Densenet results in the lowest overall error. The more costly 4D convolutions also affect inference times, which would be important for real-time tracking. While the 3D CNNs can provide position estimates with up to 166 Hz, our 4D CNNs still achieve 50 Hz. Considering that there are no optimized 4D operations available yet, these results are promising for real-time applications such as motion tracking. Overall, we provide a comprehensive study of 4D spatio-temporal CNNs in comparison to their 3D counterparts and demonstrate that position estimations of an object can be improved significantly when a stream of volumes is used. As our methods are generic, they can be easily transferred to other tasks or imaging modalities where sequences of volumetric images are of interest, e.g., motion tracking based on volumetric ultrasound or magnetic particle imaging.

\bibliographystyle{style_bib.bst}
\bibliography{Referenzen.bib}   
\end{document}